\documentclass{article}
\newcommand{\nin}{\noindent}
\newcommand{\be}{\begin{equation}}
\newcommand{\ee}{\end{equation}}
\newcommand{\bea}{\begin{eqnarray}}
\newcommand{\eea}{\end{eqnarray}}

\newcommand{\hf}{\frac{1}{2}}
\newcommand{\nn}{\nonumber\\}

\title{Non-renormalization for the Liouville wave function}

\author{J. Alexandre\footnote{jean.alexandre@kcl.ac.uk}, Anna Kostouki\footnote{anna.kostouki@kcl.ac.uk}
and N. E. Mavromatos\footnote{nikolaos.mavromatos@kcl.ac.uk}\\
Department of Physics, King's College London\\
WC2R 2LS, UK}

\begin{document}

\maketitle

\begin{abstract}

Using an exact functional method, within the framework of the gradient expansion for the Liouville
effective action, we show that the kinetic term for the Liouville field is not renormalized.

\end{abstract}

\section{Introduction}

\nin It is known that the effective potential for the two-dimensional Liouville theory remains an exponential, with renormalized coupling constant and mass parameter \cite{jackiw}. This property respects the symmetry of the
classical action, under which a translation in the Liouville field is equivalent to a change in the mass parameter.

We study here the wave function renormalization $Z$ of the Liouville field, using an exact functional method, 
which leads
to a self-consistent equation for the effective action (the proper graphs generator functional), in the spirit of a
Schwinger-Dyson equation, and which is therefore not based on a loop expansion.
The idea is to look at the evolution of the
quantum theory with the amplitude of the central charge deficit $Q^2$ of the Liouville theory \cite{ddk}, since it was shown in \cite{AEM2} that it is possible to
obtain exact flows for the quantum theory with $Q^2$. As we emphasize below,
these flows are regularized  by a {\it fixed} world sheet cut off,
unlike the Wilsonian approach. Using this method,
it was already found in \cite{AEM2}, in the approximation where $Z$ does not depend on the Liouville field, that $Z$ does not get quantum corrections and keeps its classical value.
We extend here this study to the more general situation where
$Z$ could be a polynomial of the Liouville field. This is the next step in order to have
a complete picture, consistent with the gradient
expansion. As we shall demonstrate below the result is similar to that of \cite{AEM2}: the wave function renormalization remains trivial, and the kinetic term for the
Liouville field does not get dressed by quantum fluctuations. 

We note that the functional approach used here, which serves as an alternative to Wilsonian renormalization, has proven to give new insights into the quantum structure of a theory, and led to non-trivial results
in a variety of contexts so far, including scalar field theory \cite{scalar}, Quantum Electro-Dynamics \cite{QED}, Wess-Zumino \cite{WZ} and
 Kaluza-Klein \cite{KK} models, time-dependent bosonic strings \cite{string}.

The structure of our article is the following: 
In section 2, we explain in some detail the functional method, already used in \cite{AEM2},
and derive the exact equations for the evolution of the potential and the
wave function renormalization with the central charge deficit, $Q^2$. 
The details of the derivations are given in Appendix A. We emphasize the specific r\^ole played by the \emph{two-dimensional} field theory in ensuring the wave function non-renormalization,
and we give the solution for the corresponding effective potential. In Section 3 we demonstrate the consistency of our results with the Wilsonian
approach, where we explain that this trivial solution for $Z$ is consistent with
an exact renormalization equation. Finally, in Appendix B we show the equivalence between the Wilsonian and the one-particle irreducible effective potentials.

\section{Evolution equations}

The bare action for the Liouville field, on a flat world sheet we assume in this work, reads:
\be\label{liouvsmodel}
S=\int d^2\xi\left\{\frac{Q^2}{2}\partial_a\phi\partial^a\phi+\mu^2e^\phi\right\},
\ee
where the amplitude of the kinetic term is controlled by the central charge deficit $Q^2$.
Upon quantization of this theory, as we explain below~\cite{AEM2}, $Q^2$ controls the amplitude of quantum fluctuations:
\begin{itemize}
\item for $Q^2>>1$, the quadratic kinetic term dominates the bare Lagrangian and therefore the
quantum theory is almost classical;
\item when $Q^2$ decreases, quantum fluctuations gradually appear in the system and the full quantum
theory is obtained when $Q^2\to$ finite constant.
\end{itemize}
Our motivation is to find the evolution of the proper graphs generator functional with $Q^2$, and therefore obtain
information on the quantum theory.

\subsection{Path integral quantization}

In order to define the corresponding quantum theory, one first defines the partition function (assuming a Euclidean world sheet, as required for convergence of the respective path integral)
\be
Z[j]=\int{\cal D}[\phi]\exp\left\lbrace -S-\int j\phi\right\rbrace =\exp\left\lbrace -W[j]\right\rbrace ,
\ee
where $j$ is the source and $W$ is the connected graphs generator functional. The classical field is defined as
\be\label{defphic}
\phi_c=\frac{\delta W}{\delta j},
\ee
and the proper graphs generator functional $\Gamma$, describing the quantum theory, is obtained as the
Legendre transform of $W$:
\be
\Gamma[\phi_c]=W[j]-\int d^2\xi ~j\phi_c,
\ee
where the source $j$ is to be understood as a functional of $\phi_c$, found by inverting
the definition (\ref{defphic}).
One obtains then a family of quantum theories, parametrized by $Q^2$; it was shown in \cite{AEM2}
that the effective action $\Gamma$ satisfies the following exact evolution equation with $Q^2$
(we omit the subscript $_c$ for the classical field)
\be\label{evolG}
\dot\Gamma=\hf\int d^2\xi~\partial_a\phi\partial^a\phi
+\hf\mbox{Tr}\left\{\frac{\partial}{\partial\xi_a}\frac{\partial}{\partial\zeta^a}
\left(\frac{\delta^2\Gamma}{\delta\phi_\xi\delta\phi_\zeta}\right)^{-1}\right\},
\ee
where the dot represents a derivative with respect to $Q^2$.
The evolution equation (\ref{evolG}) is exact and does not rely on any loop expansion: it is a self-consistent
equation, in the spirit of a differential Schwinger-Dyson equation.
We stress here that the trace appearing in eq.(\ref{evolG}) is regularized with a {\it fixed} world sheet cut
off $\Lambda$,
and the running parameter is $Q^2$, unlike the Wilsonian approach where, for a fixed $Q^2$, one would study
the evolution of $\Gamma$ with a running world sheet cut off.\\
In the framework of the gradient expansion, which we adopt in this work, we consider 
the projection on a specific subspace of functionals in the functional space where $\Gamma$ lives,
for which we assume the following form of the effective action
\be\label{ansatz}
\Gamma=\int d^2\xi\left\{ \frac{Z_Q(\phi)}{2}\partial_a\phi\partial^a\phi+V_Q(\phi)\right\}.
\ee
As we show in Appendix A,
the evolution equations with $Q^2$ for the potential $V_Q(\phi)$ and the wave
function renormalization $Z_Q(\phi)$ are
\bea\label{evolVZ}
\dot V&=&\frac{\Lambda^2}{8\pi Z}-\frac{V^{''}}{8\pi Z^2}
\ln\left(1+\frac{Z\Lambda^2}{V^{''}}\right) \nn
\dot Z&=&1+\frac{1}{8\pi Z}\left(\frac{Z^{'}}{Z}\right)^2
\left[ 5\ln\left(1+\frac{Z\Lambda^2}{V^{''}}\right)-\frac{47}{6}\right]\nn
&&~~~~~+\frac{7}{24\pi Z}\left( \frac{Z^{'}}{Z}\right) \left( \frac{V^{'''}}{V^{''}}\right),
\eea
where $Z=Z_Q(\phi)$ and $V=V_Q(\phi)$, and a prime denotes derivative with respect to $\phi$.

As can be seen from the evolution equations (\ref{evolVZ}), a solution
where $Z$ does not depend on the Liouville field (i.e. $Z^{'}=0$) is consistent, for which case we also obtain $\dot Z=1$, and
therefore no renormalization of the wave function.

One could seek for other solutions, different from $Z=Q^2$, but
we will give below several arguments in favour of the uniqueness of the $\phi$-independent solution:

\begin{itemize}

\item As discussed in section 3, the solution $Z=Q^2$ is consistent with an exact 
renormalization equation
for the potential, using a sharp cut off.

\item Also in the Wilsonian context, the Liouville theory has been studied using the average action formalism~\cite{reuter}, based on a smooth cut off procedure, thereby allowing the study of the evolution of the
wave function renormalization.
In this work, the wave function renormalization $Z_k(\phi)$, where $k$ is the running cut off,
does depend on the Liouville field, as
a consequence of the initial condition of the flows, which is chosen so as to satisfy the respective Weyl-Ward identities.
The authors argue, though, that the IR limit $k\to 0$ of the average action, which corresponds to the
effective action we consider here, is consistent with this non-renormalization property.

\item We can imagine integrating the equation for $Z_Q$ numerically, starting from
the initial condition $Z_Q(\phi)\simeq Q^2$ for $Q^2>>1$, since the theory is then almost classical.
The step $Q\to Q-dQ$ corresponds to $Q^2\to Q^2+dx$, with $dx=-2QdQ+dQ^2$, and we have then
$$
Z_{Q-dQ}=Z_Q+dx(\dot Z_Q)=Z_Q+dx
$$
because $Z^{'}=0$ for the initial condition. Therefore
$$
Z_{Q-dQ}=Z_Q-2QdQ+dQ^2=(Q-dQ)^2,
$$
In this way we arrive, step by step, to the result that $Z=Q^2$ for any value of $Q$.

\item We show in the next subsection that, for a field-independent $Z$,
this non-renormalization property is possible in dimension $d=2$ only, which gives a strong
indication that the solution  $Z=Q^2$ is the relevant one in the more general case 
studied here. This is also consistent with the world-sheet conformal-invariance restoring properties of the Liouville theory~\cite{ddk};

\end{itemize}

Finally, in the case of a curved world sheet, the bare action contains an additional term, linear in the Liouville field,
and reads
\be
S=\int d^2\xi\sqrt\gamma\left\lbrace \frac{Q^2}{2}\gamma_{ab}\partial^a\phi\partial^b\phi+Q^2R^{(2)}\phi+
\mu^2 e^\phi\right\rbrace,
\ee
where $\gamma_{ab}$ is the world sheet metric, with determinant $\gamma$ and curvature scalar $R^{(2)}$.
The gradient expansion for the effective action would then have to take into account this linear term in $\phi$, but
because of the
second functional derivative, which appears in the evolution equation (\ref{evolG}), this additional linear term  does not play a r\^ole in the generation of quantum fluctuations.
It is in this sense that working, from the beginning with flat world sheets, suffices for our purposes.

\subsection{Specificity of two dimensions ($d=2$)}

In this subsection we go through the same steps as those described in Appendix A, for a wave function renormalization
independent of $\phi$, but in any dimension $d$. We show then that the renormalization of the wave function renormalization
vanishes \emph{only} for the case $d=2$.

We assume that the effective action has the form
\be
\Gamma=\int d^d\xi\left\lbrace \frac{Z_Q}{2}\partial_a\phi\partial^a\phi+V_Q(\phi)\right\rbrace,
\ee
such that its second functional derivative in configuration space reads:
\be
\frac{\delta^2\Gamma}{\delta\phi_\xi\delta\phi_\zeta}=\left( -Z\partial_a\partial^a
+V^{''}\right) \delta^{(2)}(\xi-\zeta).
\ee
For the configuration $\phi=\phi_0+2\rho\cos(k\xi)$, where $\phi_0,\rho,k$ are constants,
the second functional derivative in Fourier space is:
\bea
\frac{\delta^2\Gamma}{\delta\phi_p\delta\phi_q}&=&
\left( Zp^2+V^{''}+\rho^2V^{(4)}\right) (2\pi)^2\delta^{(2)}(p+q)\nn
&&+\rho V^{(3)}(2\pi)^2\left( \delta^{(2)}(p+q+k)+\delta^{(2)}(p+q-k)\right) \nn
&&+\frac{\rho^2}{2}V^{(4)}\left( \delta^{(2)}(p+q+2k)+\delta^{(2)}(p+q-2k)\right)\nn
&&+\mbox{higher orders in $\rho$}.
\eea
The inverse of this second functional derivative is calculated using
\be
(A+B)^{-1}=A^{-1}- A^{-1}BA^{-1}+A^{-1}BA^{-1}BA^{-1}+\cdot\cdot\cdot
\ee
where $A$ stands for the diagonal contribution and $B$ for the off-diagonal one, proportional to $\rho$.
The relevant term for the evolution of $Z$ is
\bea\label{exp}
&&\mbox{Tr}\left\lbrace p^2A^{-1}BA^{-1}BA^{-1}\right\rbrace\nn
&=&{\cal A}\rho^2\left( V^{(3)}\right)^2\int\frac{d^dp}{(2\pi)^d}\frac{p^2}{f^2(p)}\left( \frac{1}{f(p+k)}+\frac{1}{f(p-k)}\right)\nn
&=&2{\cal A}\rho^2\left( V^{(3)}\right)^2I(k),
\eea
where ${\cal A}$ is the two-dimensional volume,  $f(p)=Zp^2+V^{''}$ and
\be
I(k)=\int\frac{d^dp}{(2\pi)^d}\frac{p^2}{f^2(p)f(p+k)}
\ee
The evolution of the wavefunction renormalization, $Z$, is proportional to the quadratic-order-in-$k$ part of $I(k)$, and we have
\bea\label{Ik}
I(k)
&=&I(0)+\int\frac{d^dp}{(2\pi)^d}\left\lbrace \frac{4Z^2p^2(k\cdot p)^2}{(Zp^2+V^{''})^5}
-\frac{Zk^2p^2}{(Zp^2+V^{''})^4}\right\rbrace +{\cal O}(k^4)\\
&=&I(0)+k^2Z^{-d/2}\int\frac{d^dp}{(2\pi)^d}\left\lbrace \frac{2p^4}{(p^2+V^{''})^5}
-\frac{p^2}{(p^2+V^{''})^4}\right\rbrace +{\cal O}(k^4)\nn
&=&I(0)+k^2\frac{\pi^{d/2}}{(2\pi)^d}\frac{Z^{-d/2}}{[V^{''}]^{3-d/2}}\left\lbrace
2\frac{\Gamma(3-d/2)}{\Gamma(3)}-2\frac{\Gamma(5-d/2)}{\Gamma(5)}\right. \nn
&&~~~~~~~~~~~~~~~~~~~~~
-2d\frac{\Gamma(4-d/2)}{\Gamma(5)}\left. -\frac{d}{2}\frac{\Gamma(3-d/2)}{\Gamma(4)}\right\rbrace
+{\cal O}(k^4)\nonumber
\eea
Using the property $\Gamma(n+1)=n\Gamma(n)$, together with $\Gamma(1)=1$, the expansion (\ref{Ik}) can be written
\be
I(k)=I(0)+k^2\frac{\pi^{d/2}}{(2\pi)^d}\frac{Z^{-d/2}}{[V^{''}]^{3-d/2}}\Gamma(3-d/2)\frac{d}{24}
\left( \frac{d}{2}-1\right)+{\cal O}(k^4),
\ee
which shows that the term of quadratic order in $k$ vanishes for $d=2$ only. This is a strong indication that the solution $\dot Z=1$ found previously is the relevant one.

\subsection{Solution for the potential}

From now on, we consider $Z=Q^2$. The evolution equation (\ref{evolVZ}) for the potential becomes then
\be\label{dotV}
\dot V=-\frac{V^{''}}{8\pi Q^4}
\ln\left(1+\frac{Q^2\Lambda^2}{V^{''}}\right),
\ee
where the quadratic divergence was disregarded, as it is field-independent.

The equation (\ref{dotV}) has been studied in \cite{AEM2} for the specific regimes $Q^2\to 0$ and $Q^2\to\infty$. We give here some details of the derivation for finite values of $Q^2$.  We therefore assume that
\be\label{condition}
\frac{Q^2\Lambda^2}{V^{''}}>>1.
\ee
With this condition in mind, eq.(\ref{dotV}) is then satisfied by a potential of the form
\be\label{solV}
V_Q(\phi)=\Lambda^2 v_Q\exp\left( \varepsilon_Q\phi\right) ,
\ee
where $v_Q$ and $\varepsilon_Q$ are dimensionless functions of $Q$ (for the condition
(\ref{condition}) to be satisfied we need  $v_Q<<1$). Indeed,
plugging this ansatz into the evolution equation (\ref{dotV}) gives, in the limit (\ref{condition}),
\bea
\dot v&=&-\frac{v\varepsilon^2}{8\pi Q^4}\ln\left( \frac{Q^2}{v\varepsilon^2}\right) \nn
\dot \varepsilon&=&\frac{\varepsilon^3}{8\pi Q^4}.
\eea
The latter evolution equation for $\varepsilon$ can be integrated exactly. The appropriate boundary condition is $\varepsilon\to 1$ when $Q^2\to\infty$, since the system is then classical.
The integration over $Q^2$ leads to
\be\label{solepsilon}
\varepsilon_Q=\sqrt\frac{4\pi Q^2}{1+4\pi Q^2}.
\ee
We remind the reader that the solution (\ref{solepsilon}) is {\it exact} in the framework of the gradient
expansion (\ref{ansatz}), and is not based on a loop expansion.
The evolution equation for $v_Q$ is not solvable exactly, and we thus leave the study of the
potential amplitude for the next section, where this is achieved by means of a Wilsonian exact renormalization approach.

Before closing this section, we note that, for the specific conformal charge deficit $Q^2=8$, corresponding to $c=1$ conformal field theories, there are two cosmological constant operators, dressing the identity
$(\mu_1^2+\mu_2^2\phi)\exp(\sqrt{2}\phi)$ \cite{moore}, where $\mu_1,\mu_2$ are constants. Our solution above cannot include the 
operator proportional to $\mu_2^2$ \cite{AEM2}, since we consider a continous set of values for $Q^2$ and 
this operator exists only for a 
discrete isolated value. To incorporate this case, one should study the flow with respect to 
another parameter in the bare theory with fixed $Q^2=8$, such as $\alpha^{'}$ or $\mu_i$.

\section{Consistency with the Wilsonian picture}

We now exhibit the Wilsonian properties of the solution (\ref{solV}),
using the exact renormalization method of \cite{WH}.
We consider an initial two-dimensional bare theory, with running cut-off $\Lambda$. The
effective theory defined at the scale $\Lambda-\delta\Lambda$ is derived by integrating the ultraviolet
degrees of freedom from $\Lambda$ to $\Lambda-\delta\Lambda$. The idea of exact renormalization methods is to
perform this integration infinitesimally, i.e. take the limit $\delta\Lambda/\Lambda\to 0$,
which leads to an exact evolution equation for $S_\Lambda$.
The procedure was detailed in \cite{WH}, and here we reproduce
only the main steps for clarity and completeness. Note that we consider here a sharp cut-off,
which is possible only if we consider 
the evolution of the potential part of the Wilsonian action, as explained now.

We consider a Euclidean two-dimensional spacetime, and we assume that,
for each value of the energy scale $\Lambda$, the Euclidean action $S_\Lambda$ has the form
\be\label{ansatzbis}
S_\Lambda=\int d^2\xi\left\lbrace \frac{Z_\Lambda(\phi)}{2}\partial_a\phi\partial^a\phi+V_\Lambda(\phi)\right\rbrace .
\ee
The integration of the ultraviolet degrees of freedom is implemented in the following way. We write the
dynamical fields $\phi=\phi_{IR}+\psi$, where the $\phi_{IR}$ is the infrared field with
non-vanishing Fourier components
for $|p|\le \Lambda-\delta\Lambda$, and $\psi$ is the degree of freedom to be integrated out, with
non-vanishing Fourier components for $\Lambda-\delta\Lambda<|p|\le \Lambda$ only.
An infinitesimal step of the renormalization group transformation reads:
\bea\label{transfo}
&&\exp\left(-S_{\Lambda-\delta\Lambda}[\phi_{IR}]+S_\Lambda[\phi_{IR}]\right)\\
&=&\exp\left(S_\Lambda[\phi_{IR}]\right)\int {\cal D}[\psi]\exp\left(-S_\Lambda[\phi_{IR}+\psi]\right)\nn
&=&\int{\cal D}[\psi]\exp\left(-\int_\Lambda \frac{\delta S_\Lambda[\phi_{IR}]}{\delta\psi(p)}\psi(p)
-\hf\int_\Lambda\int_\Lambda\frac{\delta^2S_\Lambda[\phi_{IR}]}{\delta\psi(p)\delta\psi(q)}\psi(p)\psi(q)\right),\nn
&&~~~~~~~~~~~~~~+\mbox{higher orders in}~\delta\Lambda, \nonumber
\eea
where $\int_\Lambda$ represents the integration over Fourier modes for $\Lambda-\delta\Lambda<|p|\le \Lambda$.
Higher-order terms in the expansion of the action are indeed of higher order in $\delta\Lambda$, since each integral
involves a new factor of $\delta\Lambda$. The only relevant terms are of first and second order in
$\delta\Lambda$ \cite{WH}, which are at most quadratic in the dynamical
variable $\psi$, and therefore lead to a Gaussian integral. We then have
\bea\label{evolS}
&&\frac{S_\Lambda[\phi_{IR}]-S_{\Lambda-\delta\Lambda}[\phi_{IR}]}{\delta\Lambda}\nn
&=&\frac{\mbox{Tr}_\Lambda}{\delta\Lambda}\left\{\frac{\delta S_\Lambda[\phi_{IR}]}{\delta\psi(p)}
\left(\frac{\delta^2S_\Lambda[\phi_{IR}]}{\delta\psi(p)\delta\psi(q)}\right)^{-1}
\frac{\delta S_\Lambda[\phi_{IR}]}{\delta\psi(q)}\right\}\nn
&&-\frac{\mbox{Tr}_\Lambda}{2\delta\Lambda}\left\{\ln\left(\frac{\delta^2S_\Lambda[\phi_{IR}]}{\delta\psi(p)\delta \psi(q)}\right)\right\}
+{\cal O}(\delta\Lambda),
\eea
where the trace Tr$_\Lambda$ is to be taken in the shell of thickness $\delta\Lambda$, and is therefore
proportional to $\delta\Lambda$.

We are interested in the evolution equation for the potential only, for which it is sufficient to
consider a constant infrared configuration $\phi_{IR}=\phi_0$, and  this is the reason why a sharp cut-off can be used:
the singular terms that could arise from the $\theta$ function, characterizing the sharp cut-off,
are not present, since the derivatives of the infrared field vanish.
In this situation, the first term on the right-hand
side of eq.(\ref{evolS}), which
is a tree-level term, does not contribute: $\delta S_\Lambda/\delta\psi(p)$ is proportional
to $\delta^2(p)$, and thus has no overlap with the domain of integration $|p|=\Lambda$.
We are therefore left with the second term, which arises from quantum fluctuations, and
the limit $\delta\Lambda\to 0$ gives, with the ansatz (\ref{ansatzbis}),
\be\label{evolpot}
\partial_\Lambda V_\Lambda(\phi_0)-\partial_\Lambda V_\Lambda(0)=
-\frac{\Lambda}{4\pi}\ln\left(\frac{Z_\Lambda(\phi_0)\Lambda^2+V^{''}_\Lambda(\phi_0)}
{Z_\Lambda(0)\Lambda^2+V^{''}_\Lambda(0)}\right),
\ee
Eq.(\ref{evolpot}) provides a resummation of all the loop orders, since it consists in a self-consistent
equation. As a result, the
evolution equation (\ref{evolpot}) is exact within the framework of the ansatz (\ref{ansatzbis}),
and is independent of a loop expansion.

In order to make the connection with the solution (\ref{solV}), we now consider the following ansatz
\bea\label{solL}
Z_\Lambda(\phi_0)&=&Q^2\\
V_\Lambda(\phi_0)&=&\Lambda^2v_\Lambda\exp(\varepsilon \phi_0)\nonumber,
\eea
where $\varepsilon$ is the constant (\ref{solepsilon}) and $v_\Lambda$ depends on the running cut off only.
One should keep in mind here that $Q^2$ is now {\it constant}, whereas the cut off $\Lambda$ is {\it running}.
When plugged in the Wegner-Houghton equation (\ref{evolpot}), the ansatz (\ref{solL}) leads to
\be\label{agaga}
\left( 2\Lambda v+\Lambda^2\partial_\Lambda v\right) \left( \exp\left( \varepsilon\phi_0\right) -1\right)
=-\frac{\Lambda}{4\pi}\ln\left( \frac{Q^2+\varepsilon^2 v\exp(\varepsilon\phi_0)}{Q^2+\varepsilon^2 v}\right).
\ee
One can see that
this equation is consistent in the limit $v<<1$ only, which we are interested in:
keeping the first
orders in $v$, the $\phi_0$-dependence cancels out and the remaining equation is
\be
2\Lambda v+\Lambda^2\partial_\Lambda v=-\frac{\Lambda}{4\pi Q^2}\varepsilon^2 v,
\ee
which is easily integrated to
\be
v_\Lambda=\left( \frac{\mu}{\Lambda}\right)^{2+\varepsilon^2/(4\pi Q^2)}.
\ee
This solution indeed satisfies $v<<1$, since we are interested in the regime of large cut off, in the spirit of
the condition (\ref{condition}).
Taking into account the solution (\ref{solepsilon}), the potential is finally
\be\label{finalsol}
V_\Lambda(\phi)=\mu^2\left( \frac{\mu}{\Lambda}\right)^{1/(1+4\pi Q^2)}
\exp\left(\phi\sqrt\frac{4\pi Q^2}{1+4\pi Q^2}\right) ,
\ee
We stress again that this solution is not the result of a loop expansion.
An important remark is in order here: it was possible to find the solution (\ref{finalsol})
of the Wilsonian exact renormalization group equation, {\it because} $Z$ does not depend on $\phi_0$.
Indeed, it is the only possibility for the $\phi_0$-dependence to cancel in eq.(\ref{agaga}), at the first order in $v$.
This shows the consistency of the choice  $Z=Q^2$ made in the previous section.

Note that the solution (\ref{finalsol}) does not need to satisfy the evolution equation (\ref{dotV}),
since the Wilsonian potential defined in this section is not obtained by means of the Legendre transform
as the potential defined in the previous section. The equivalence between these potentials is obtained
in the limit where the running cut off goes to 0 (see Appendix B),
but in this case, the expression (\ref{finalsol}) is not
valid since it was derived in the limit of large cut off.

Finally, the limit of the Wilsonian potential (\ref{finalsol}) when $Q^2\to\infty$, for fixed cut off (in the spirit of
section 2), gives the expected bare Liouville potential shown in eq.(\ref{liouvsmodel}).

\section{Discussion}

In this work we have analysed Liouville field theory on the world-sheet from the perspective of a novel functional method, suggested in \cite{AEM2}. In particular, we have demonstrated that the function $Z(\phi)$ appearing in the Liouville kinetic term, is not renormalized, that is it preserves its classical form $Z=Q^2$ in the full quantum theory.
As we have discussed, this is a specific feature of the two-dimensional field theory, and is not the case in general, e.g. in four dimensions~\cite{scalar}.
In fact, this feature is also essential for maintaining the conformal properties of the Liouville field, in particular its r\^ole in restoring conformal symmetry \cite{ddk}.

It should be stressed that in the present work we have assumed a functional dependence of the effective theory based on the gradient expansion, namely that $Z(\phi)$ is only \emph{a polynomial} function of the Liouville field $\phi$ and \emph{not} its world-sheet derivatives $\partial \phi$. This assumption is dictated by the above-mentioned argument of conformal covariance of the Liouville action, which we wish to maintain in the full quantum theory \cite{ddk}. It is remarked that in case one included such higher derivative terms, the allowed structures in the effective action should involve terms of the form
\begin{equation}\label{str}
Z_n(\phi) \partial_a \phi \left(\frac{\partial_b\partial^b}{\mu^2}\right)^n\partial^a \phi
\end{equation}
where $Z_n(\phi)$ are dimensionless polynomials of $\phi$ and $n$ is an integer.
It remains to be seen whether in such cases the above-mentioned non renormalization result is valid. However,
we expect such terms not to be present, as their presence would appear in conflict with the standard
conformal properties of the Liouville field. In this sense we think
that the analysis in the present article is complete.

As a final remark we note that for a curved world sheet, with a curvature scalar $R^{(2)}$, replacing $\mu^2$ in (\ref{str}), one could in principle have structures of the form
\begin{equation}\label{yn}
Y_n(\phi) \partial_a \phi \left(\frac{\partial_b\partial^b}{R^{(2)}}\right)^n\partial^a \phi
\end{equation}
where $Y_n(\phi)$ are dimensionless polynomials of $\phi$. However, such structures cannot appear for $n \ne 0$, as they should vanish in the limit of flat world sheet $R^{(2)}\to 0$, and since $Y_n$ does not depend on
the curvature scalar, it cannot vanish in this limit in order to leave the terms (\ref{yn}) finite.

Before closing we note that the above analysis can be extended to incorporate Liouville-dressed non-critical stringy $\sigma$-models, involving the coupling of $X^\mu$ fields with the Liouville mode $\phi$. In such a case there are more complicated potential terms, since for each non conformal vertex operator $V(X)$ of the non-critical string, there is a conformal-symmetry restoring factor $e^{\alpha \phi}$, with $\alpha$ the appropriate Liouville dimension~\cite{ddk}, multiplying $V(X)$, $\int d^2 \sigma e^{\alpha \phi}V(X)$. Nevertheless, the application of the exact method for the Liouville sector and the associate $Q^2$ flows applies to this case, with similar results, as far as the Liouville wavefunction non renormalization is concerned.

\section*{Acknowledgements}

J. A. would like to thank Janos Polonyi for useful discussions related to the flow equations,
and the explanation given in Appendix B.
We wish to thank the organisers of the \emph{1st Annual School Of EU Network ``UniverseNet, The Origin Of The Universe: Seeking Links Between Fundamental Physics And Cosmology}, Mytilene (Island of Lesvos, Greece), September 24-29 2007, for the hospitality and for giving the opportunity to A.K. to present preliminary results of this work.
The work of of A.K. and N.E.M is partially supported by the European Union through the FP6 Marie Curie Research and Training Network {\it UniverseNet} (MRTN-CT-2006-035863).

\section*{Appendix A: derivation of the flow equations}

With the following ansatz for the effective action,
\be
\Gamma=\int d^2\xi\left\lbrace \frac{Z_Q(\phi)}{2}\partial_a\phi\partial^a\phi+V_Q(\phi)\right\rbrace ,
\ee
the second functional derivative appearing in the evolution equation (\ref{evolG}) is
\bea
\frac{\delta^2\Gamma}{\delta\phi_\xi\delta\phi_\zeta}&=&\left( -Z\partial_a\partial^a
-\frac{Z^{''}}{2}\partial_a\phi\partial^a\phi\right) \delta^{(2)}(\xi-\zeta)\nn
&&+\left( -Z^{'}\partial_a\partial^a\phi-Z^{'}\partial_a\phi\partial^a+V^{''}\right) \delta^{(2)}(\xi-\zeta).
\eea
In order to derive the evolution equation for the potential $V$, a constant configuration $\phi=\phi_0$ is sufficient. However, 
since we are also interested in the evolution of $Z$, we need a coordinate-dependent configuration, and thus 
we consider 
\be\label{phirho}
\phi(\xi)=\phi_0+2\rho\cos(k\xi),
\ee
where $\rho$ and $k$ are constants. The evolution equation for the potential
$V$ will then be obtained by identifying the terms independent of $k$ in eq.(\ref{evolG}).
On the other hand, the evolution equation
for $Z$ is obtained by identifying the terms proportional to $\rho^2 k^2$. \\
For the configuration (\ref{phirho}), the left hand side of eq.(\ref{evolG}) reads
\be\label{lhside}
\dot\Gamma={\cal A}\left\lbrace \dot V+\rho^2\dot V^{''}+\rho^2 k^2\dot Z+\cdot\cdot\cdot\right\rbrace ,
\ee
where ${\cal A}$ is the world sheet surface area, and
higher orders in $\rho$ were not written explicitly. The second derivative of $\Gamma$, in Fourier
components and to order $\rho^2$, is
\bea
\frac{\delta^2\Gamma}{\delta\phi_p\delta\phi_q}&=&A_{pq}+B_{pq},~~~~\mbox{with}\\
A_{pq}&=&\left\lbrace p^2Z+V^{''}+\rho^2\left[(p^2+k^2)Z^{''}+V^{(4)}\right] \right\rbrace (2\pi)^2\delta^{(2)}(p+q)\nn
B_{pq}&=&~~\rho\left[(p^2+kp+k^2)Z^{'}+V^{(3)}\right](2\pi)^2\delta^{(2)}(p+q+k)\nn
&&+\rho\left[(p^2-kp+k^2)Z^{'}+V^{(3)}\right](2\pi)^2\delta^{(2)}(p+q-k)\nn
&&+\frac{\rho^2}{2}\left[ (p^2+2kp+3k^2)Z^{''}+V^{(4)}\right](2\pi)^2\delta^{(2)}(p+q+2k)\nn
&&+\frac{\rho^2}{2}\left[ (p^2-2kp+3k^2)Z^{''}+V^{(4)}\right](2\pi)^2\delta^{(2)}(p+q-2k)\nonumber.
\eea
We see that $A$ is diagonal in Fourier space, whereas $B$ is not.
The inverse is then expanded in powers of $\rho$, using
\be
\left( \frac{\delta^2\Gamma}{\delta\phi\delta\phi}\right)_{pq}^{-1}=
\left( A^{-1}\right)_{pq}-\left( A^{-1}BA^{-1}\right) _{pq}+\left( A^{-1}BA^{-1}BA^{-1}\right)_{pq}+\cdot\cdot\cdot
\ee
which finally gives, to order $\rho^2$,
\bea
&&\mbox{Tr}\left\lbrace \partial_a\partial^a\left( \frac{\delta^2\Gamma}{\delta\phi\delta\phi}\right)^{-1}\right\rbrace \\
&=&{\cal A}\left\lbrace \frac{\Lambda^2}{8\pi Z}-\frac{V^{''}}{8\pi Z^2}
\ln\left(1+ \frac{Z\Lambda^2}{V^{''}}\right) \right\rbrace +\rho^2{\cal A}\left\lbrace\cdot\cdot\cdot \right\rbrace \nn
&&+\rho^2k^2{\cal A}\left\lbrace \frac{5}{4\pi Z}\left( \frac{Z^{'}}{Z}\right)^2\ln\left(1+\frac{Z\Lambda^2}{V^{''}}\right)
-\frac{47}{24\pi Z}\left( \frac{Z^{'}}{Z}\right)^2+\frac{7}{12\pi Z}\frac{Z^{'}}{Z}\frac{V^{(3)}}{V{''}}\right\rbrace
\nonumber
\eea
In the above expression, the term proportional to $\rho^2$ and independent of $k$ is not relevant, as it leads to the
evolution equation for $V^{''}$. We checked, though, that this evolution is consistent with the one for
$V$. The evolution equations for $V$ and $Z$ are then those given in eqs.(\ref{evolVZ}), after
identification with the left hand side of (\ref{lhside}).

\section*{Appendix B: Equivalence between Wilsonian and one-particle-irreducible effective potentials}

For a constant IR configuration $\phi_0$, the Wilsonian effective potential $U_{Wils}$ is defined by
\be
\exp\left( iVU_{Wils}(\phi_0)\right) =
\int{\cal D}[\phi]\exp\left( iS[\phi_0+\phi]\right),
\ee
where $V$ is the volume of space time, $S$ is the  bare action defined at some cut off $\Lambda$, and the
dynamical variable $\phi$ which is integrated out has non-vanishing Fourier components for $|p|\le\Lambda$.
One can also write the previous definition as
\bea\label{delta}
\exp\left( iVU_{Wils}(\phi_0)\right) &=&
\int{\cal D}[\phi]\exp\left( iS[\phi]\right)\delta\left( \int_x\phi-V\phi_0\right) \\
&=&\int{\cal D}[\phi]\exp\left( iS[\phi]\right)\int_j\exp\left( ij\int_x(\phi-\phi_0)\right),\nonumber
\eea
where $\int_x$ denotes the integration over space time,
$j$ is a real variable, and $\int_j$ denotes the integration over $j$.
Using the notations of section 2, this expression can be written as 
\bea
\exp\left( iVU_{Wils}(\phi_0)\right) &=&
\int_jZ[j]\exp\left( -i\int_x j\phi_0\right) \nn
&=&\int_j \exp\left(iW[j]-i\int_x j\phi_0\right)\nn
&=& \int_j \exp\left(i\Gamma[\phi_0]\right) \nn
&=& \int_j \exp\left(iVU_{1PI}(\phi_0)\right),
\eea
where $U_{1PI}$ is the one-particle-irreducible effective potential.
Note that $j$ plays the r\^ole of a constant source for the field $\phi$, leading to
the constant classical field $\phi_0$. In the last expression, the integration
over $j$ leads to a multiplicative
constant, as $\phi_0$ is fixed. Disregarding the $\phi_0$-independent terms, we then obtain 
\be
U_{Wils}(\phi_0)=U_{1PI}(\phi_0).
\ee
Note that for the above argument to be valid it is essential that we work in Minkowski space time, since the delta function in eq.(\ref{delta}) is expressed in terms of its Fourier transform.

\end{document}